\documentclass{iopart}
\expandafter\let\csname equation*\endcsname\relax % to allow for amsmath
\expandafter\let\csname endequation*\endcsname\relax

\usepackage{iopams}

\usepackage{citesort}

\usepackage{graphicx} % needed for figures
\usepackage{caption}
\usepackage{float} % locate figures, [H]
\usepackage{bm,color}

\usepackage{amssymb}
\usepackage{amsmath}

\usepackage[normalem]{ulem} %makes underlining possible: \uline \uwave \sout
\usepackage{color}%makes \textcolor{grey}{text} available
\begin{document}
\title{Robust and Ultrafast State Preparation by Ramping Artificial Gauge Potentials}

\author{Botao Wang}
\ead{botao.wang@tu-berlin.de}
\address{Institut f\"ur Theoretische Physik, Technische Universit\"at Berlin, Hardenbergstra{\ss}e 36, 10623 Berlin, Germany}
\address{Max-Planck-Institut f{\"u}r Physik komplexer Systeme, N{\"o}thnitzer Stra\ss e 38, 01187 Dresden, Germany}
\author{Xiao-Yu Dong}
\ead{xiaoyu.dong@ugent.be}
\address{Department of Physics and Astronomy, Ghent University, Krijgslaan 281, 9000 Gent, Belgium}
\address{Department of Physics and Astronomy, California State University, Northridge, 91330 CA, USA}
\address{Max-Planck-Institut f{\"u}r Physik komplexer Systeme, N{\"o}thnitzer Stra\ss e 38, 01187 Dresden, Germany}
\author{F. Nur \"{U}nal}
\ead{fnu20@cam.ac.uk}
\address{TCM Group, Cavendish Laboratory, University of Cambridge, JJ Thomson Avenue, Cambridge CB3 0HE, UK}
\address{Max-Planck-Institut f{\"u}r Physik komplexer Systeme, N{\"o}thnitzer Stra\ss e 38, 01187 Dresden, Germany}

\author{Andr\'{e} Eckardt}
\ead{eckardt@tu-berlin.de}
\address{Institut f\"ur Theoretische Physik, Technische Universit\"at Berlin, Hardenbergstra{\ss}e 36, 10623 Berlin, Germany}
\address{Max-Planck-Institut f{\"u}r Physik komplexer Systeme, N{\"o}thnitzer Stra\ss e 38, 01187 Dresden, Germany}

\vspace{10pt}

\begin{abstract}
The implementation of static artificial magnetic fields in ultracold atomic systems has become a powerful tool, e.g.\ for simulating quantum-Hall physics with charge-neutral atoms. Taking an interacting bosonic flux ladder as a minimal model, we investigate protocols for adiabatic state preparation via magnetic flux ramps. Considering the fact that it is actually the artificial vector potential (in the form of Peierls phases) that can be experimentally engineered in optical lattices, rather than the magnetic field, we find that the time required for adiabatic state preparation dramatically depends on which pattern of Peierls phases is used. This can be understood intuitively by noting that different patterns of time-dependent Peierls phases that all give rise to the same magnetic field ramp, generally lead to different artificial electric fields during the ramp. As an intriguing result, we find that an optimal choice allows for preparing the ground state almost instantaneously in the non-interacting system, which can be related to the concept of counterdiabatic driving. Remarkably, we find extremely short preparation times also in the strongly-interacting regime. Our findings open new possibilities for robust state preparation in atomic quantum simulators.
\end{abstract}

%
% Uncomment for keywords
\vspace{2pc}
\noindent{\it Keywords}: Artificial gauge fields, adiabatic preparation, optical lattices, cold atoms
%
% Uncomment for Submitted to journal title message
\submitto{\NJP}
%
% Uncomment if a separate title page is required
\maketitle
% 
% For two-column output uncomment the next line and choose [10pt] rather than [12pt] in the \documentclass declaration
%\ioptwocol
%

\section{Introduction}
The engineering of artificial magnetic fields for charge-neutral atoms in optical lattices has been a powerful tool to simulate lattice models with exotic phases including quantum Hall states and topological insulators~\cite{2011Dalibard,2013Galitski,2014Goldman,2016Goldman,2017Eckardt,2017Aidelsburger,2019Cooper}. 
More precisely, in these experiments a \textit{static} artificial gauge potential (in the form of Peierls phases) is engineered in a particular choice of gauge (relative to the plain lattice without magnetic field). Typically, this choice is made based on experimental convenience. 
For a \textit{dynamic} process, however, where these artificial gauge potentials are varied in time, this choice does not simply correspond to a gauge freedom anymore. 
This is because their temporal change generates an artificial electric field.
After initial confirmation in a trapped quantum gas~\cite{2011Lin_E}, such artificial electric forces were observed also in optical lattices~\cite{2012Struck,2013Beeler} and predicted to lead to `gauge-dependent' time-of-flight images of Bose Einstein condensates~\cite{2015Kennedy,2015LeBlanc,2010Moeller}.
More recently, theoretical investigations showed that the engineering of time-dependent artificial gauge potentials can be employed for quantized charge pumping along tailored paths in two dimensional (fractional) Chern insulators~\cite{2018Wang,2018Raviunas} and for determining the dynamics of a wave packet in synthetic dimensions~\cite{2018Yilmaz} and nonlinear systems~\cite{2020Lelas}.
With the recent advances in quantum gas microscope techniques~\cite{2009Bakr,2016Yamamoto,2016Ott_rev,2016Kuhr,2016Zupancic,2016Cocchi,2017Drewes,2017Tai}, it becomes more and more important to explore the possibilities of controlling artificial gauge potentials in both space and time. 
In this paper, we show that this technique can be exploited for the optimization of adiabatic state preparation.
Robust adiabatic state preparation is a prerequisite for the experimental investigation (quantum simulation) of interesting states of matter with atomic quantum gases.

As minimal lattice systems with artificial magnetic fields, flux ladders have recently drawn tremendous attention, including the experimental observation of chiral edge currents~\cite{2016Livi,2017An,2017Tai,2015Mancini_edge,2015Stuhl,2014Atala}, the theoretical exploration of rich phase diagrams~\cite{2001Orignac,2005Granato,2012Dhar,2013Dhar,2013Petrescu,2014Huegel,2015Greschner_spont,2016Bilitewski,2016Greschner,2020Buser,2014Wei,2015Uchino,2016Uchino,2015Piraud,2014Tokuno,2015Keles,2015Dio,2015Dio_persist,2015Natu,2016Orignac,2017Orignac,2017Sachdeva,2018Citro,2018Romen}, the investigation of Laughlin-like states~\cite{2014Grusdt,2015Petrescu,2015Cornfeld,2017Greschner,2017Strinati,2019Strinati}, the study of Hall effect~\cite{1999Prelovsek,2000Zotos,2019Greschner,2019Filippone,2021Buser} and other aspects~\cite{2016Wu,2017Kolovsky,2017Zheng,2018Filippone,2018Strinati,2019Kamar,2019Buser}.
In this work, we investigate the adiabatic preparation of the ground state in such ladder systems via continuously ramping up the corresponding Peierls phases. 
Comparing results for different patterns of Peierls phases, all giving rise to the same magnetic flux, we find that the degree of adiabaticity dramatically depends on this choice. 
As an intriguing result, the optimal choice of Peierls phases allows for an almost instantaneous preparation of the ground state. We show that for vanishing interactions, this effect can be related to counterdiabatic driving~\cite{2003Demirplak,2005Demirplak,2009Berry,2010Chen,2013Torrontegui,2019Odelin}.
However, remarkably our approach works also for very strong interactions, where a simple explanation in terms of counterdiabatic driving is not possible.

\begin{figure}
	\centering \includegraphics[width=0.85\linewidth]{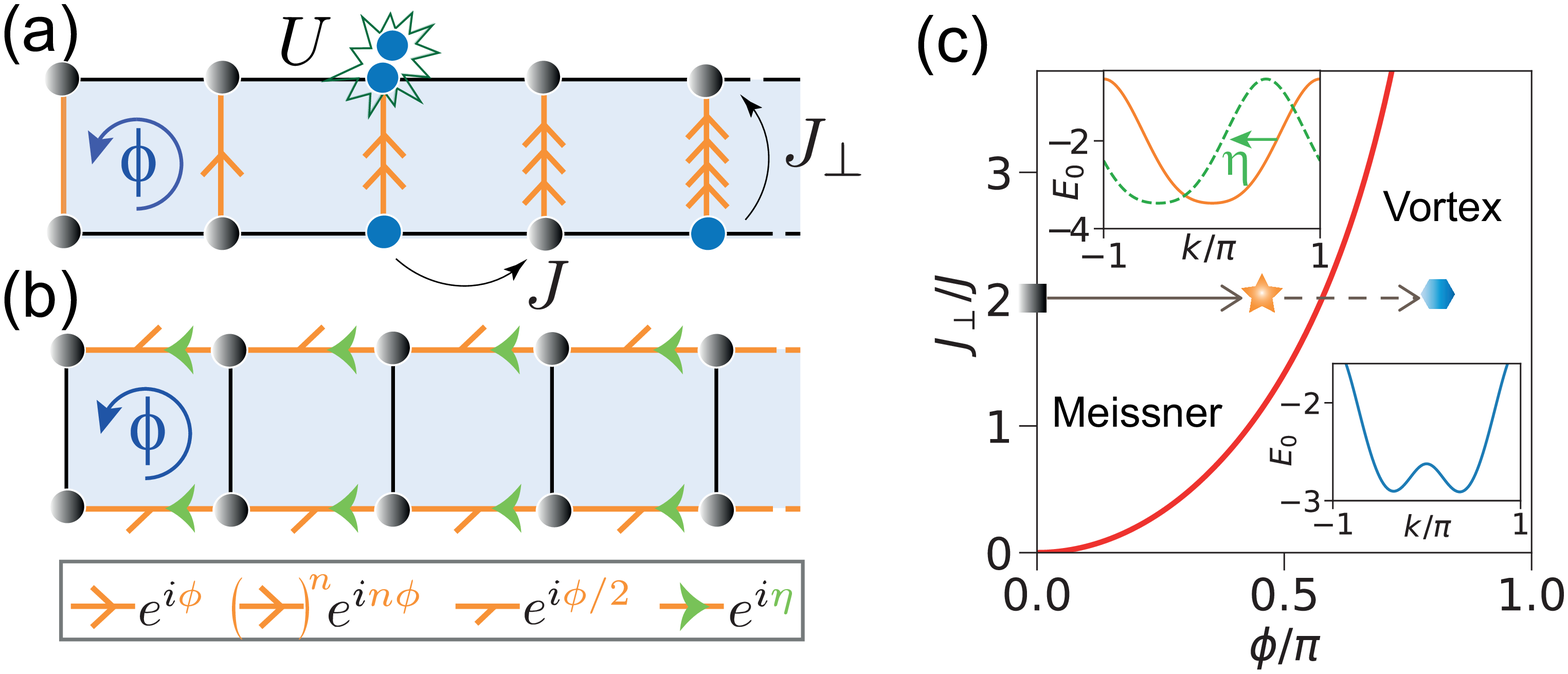}
	\caption{Bose-Hubbard ladder, with interaction parameter $U$, tunneling amplitudes $J$ ($J_\perp$) along the legs (rungs) as well as Peierls phases ${\theta_{\ell\ell'}}$ either along rungs (a) or legs (b). ${\theta_{\ell\ell'}}$ are symbolized by arrows and describe a uniform plaquette flux $\phi$. (c) Phase diagram for non-interacting system. Upper inset shows the lowest Bloch band with single minimum in the Meissner phase, for $J_\perp=2$ and Peierls phases $\theta_{\ell'\ell}^\parallel(\phi=\pi/2,\eta)$ with $\eta=0$ (solid orange line) and $\eta=\pi/2$ (dashed green line). Lower inset shows double minima of the lowest band in the vortex phase with $\phi=4\pi/5$ and $\eta=0$. The horizontal arrows indicate the paths for our state preparation via ramping artificial magnetic flux.}
	\label{fig_scheme}
\end{figure}

\section{Model}
We consider interacting bosons in a two-leg ladder described by the Bose Hubbard model
\begin{equation}
\hat{H}=-{\displaystyle \sum_{\left\langle \ell,\ell^{\prime}\right\rangle }}J_{\ell^{\prime}\ell}e^{i\theta_{\ell^{\prime}\ell}}\hat{a}_{\ell^{\prime}}^{\dagger}\hat{a}_{\ell}+\frac{U}{2}{\displaystyle \sum_{\ell}\hat{n}_{\ell}(\hat{n}_{\ell}-1)},
\end{equation}
with bosonic creation operator $\hat{a}_{\ell}^{\dagger}$ and number operator $\hat{n}_\ell=\hat{a}^{\dag}_{\ell}\hat{a}_{\ell}$ on site $\ell$. The nearest-neighbor tunneling amplitude $J_{\ell^{\prime}\ell}$ equals $J$ along legs and $J_{\perp}$ along rungs, and it is accompanied by the Peierls phase $\theta_{\ell^{\prime}\ell}$. $U$ is the on-site repulsive interaction energy. In the following, we use $J$, $\hbar/J$ and lattice constant $a$ as units for energy, time and lengths, respectively.

Due to the complex tunneling matrix elements, the accumulated net phase around one lattice plaquette is analogous to the Aharonov-Bohm phase experienced by a charged particle in a real magnetic field. Thus the Peierls phase $\theta_{\ell^{\prime}\ell}$ plays the role of a vector potential, and each set of time-independent Peierls phases $\{\theta_{\ell^{\prime}\ell}\}$ that gives the same plaquette flux reflects a gauge choice. 
A uniform flux $\phi$ can be realized, for instance, by using gauge potentials along rungs, $\theta_{\ell'\ell}^\perp(\phi)$~[Fig.~\ref{fig_scheme}(a)], or along legs, $\theta_{\ell'\ell}^\parallel(\phi,\eta)$~[Fig.~\ref{fig_scheme}(b)], with the phase $\eta$ describing a continuous family of Peierls phases.
However, when $\phi$ and $\eta$ vary in time, $\theta_{\ell'\ell}^\perp(\phi)$ and $\theta_{\ell'\ell}^\parallel(\phi,\eta)$ no longer describe gauge choices, but different artificial electric fields.

\section{Non-interacting case}
Let us start with the non-interacting limit $(U=0)$, for which the phase diagram is shown in Fig.~\ref{fig_scheme}(c). 
For weak magnetic flux, the dispersion relation of the lowest band possesses a unique minimum and the ground state exhibits currents along the leg, resembling the screening currents of the Meissner phase (MP) of a superconductor. Increasing the flux beyond the phase boundary defined by $J_\perp=2\sin(\phi/2)\tan(\phi/2)$, the minimum of the dispersion relation splits into two minima and rung-currents appear in the ground state allowing the formation of vortices analoguous to the vortex phase of a type-II superconductor~\cite{2014Atala,2014Huegel}.

\begin{figure}
	\centering\includegraphics[width=0.95\linewidth]{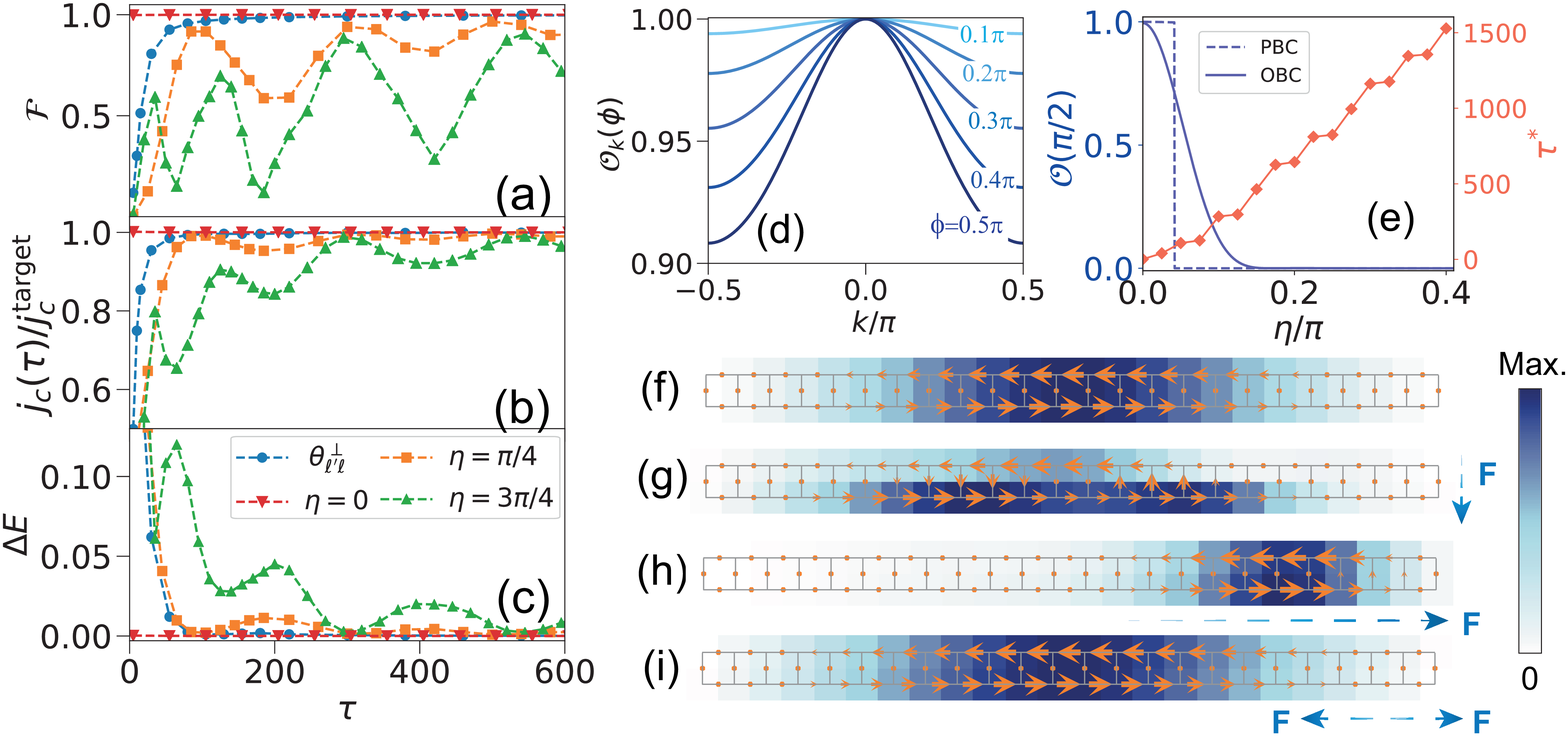}
	\caption{(a)~Fidelity, (b)~scaled chiral current and (c)~excitation energy as functions of total ramping time $\tau$ for different choices of Peierls phase configurations. The legends in (a) and (b) are the same as those in (c). (d)~The overlap $\mathcal{O}_{k}(\phi)$ for different values of $\phi$ for $\eta=0$. (e) Ground state overlap $\mathcal{O}(\pi/2)$ and minimal ramping time $\tau^{*}$ required to reach $\mathcal{F}=0.9$ as a function of $\eta$ for open (solid) and periodic (dashed) boundary conditions. Distributions of spatial density and local currents of (f) the target state, and the evolved states for (g) $\theta_{\ell'\ell}^\perp(\phi)$, $\tau=15$, (h) $\theta_{\ell'\ell}^\parallel(\phi,\eta=\pi/4)$, $\tau=200$ and (i) $\theta_{\ell'\ell}^\parallel(\phi,\eta=0)$, $\tau=1$. The darker color indicate higher densities and the size of the orange arrows along the bonds is proportional to the amplitude of probability currents. The dashed arrows F indicate directions of the average artificial electric forces.}
	\label{fig_single}
\end{figure}

In order to study adiabatic state preparation, we take our initial state and target state as the ground states of the Hamiltonian with flux $\phi=0$ and $\phi=\pi/2$, denoted as $\left|\psi_0\right\rangle$ and $\left|\psi_{\pi/2}\right\rangle$, respectively. The tunneling amplitude along rungs is fixed at $J_{\perp}=2$ so that the target state lies in the MP, as is marked by the star in Fig.~\ref{fig_scheme}(c). 
By linearly ramping the Peierls phases from zero to final values given by either $\theta_{\ell'\ell}^\perp(\phi)$ or $\theta_{\ell'\ell}^\parallel(\phi,\eta)$, the flux is continuously increased from $0$ to $\pi/2$ within the ramping time $\tau$. The evolved state $ \left|\psi(\tau)\right\rangle $ is obtained by numerically solving the Schr\"{o}dinger equation of the Hamiltonian for a finite system with $M=24$ rungs under open boundary condition.  

To quantify the degree of adiabaticity, we define the fidelity as the squared overlap between the evolved state and the target state,~$\mathcal{F}=\left|\left\langle \psi_{\pi/2}|\psi(\tau)\right\rangle \right|^2$. 
Fig.~\ref{fig_single}(a) shows the fidelities calculated by choosing artificial gauge potentials $\theta_{\ell'\ell}^\perp(\phi)$ and $\theta_{\ell'\ell}^\parallel(\phi,\eta)$ with $\eta=\{0,\pi/4,3\pi/4\}$~[cf.~legend in Fig. 2(c)].
For gauge potentials on the rungs, we find fidelities close to 1 for ramping times on the order of $\tau=300$. For gauge potentials on the legs, this time scale strongly depends on $\eta$. Remarkably, it vanishes in the limit of $\eta=0$, so that the ground state can be prepared by switching on the gauge potentials abruptly. This picture is confirmed also by looking at two other quantities characterizing the evolved state.
One is the chiral current $j_c(\tau)$ scaled by its target value $j_c^{\text{target}}$~[Fig.~\ref{fig_single}(b)], which can be readily measured in experiment~\cite{2014Atala,2015Stuhl,2015Mancini_edge} and which plays a key role in charactering different phases in a ladder system~\cite{1991Silva,2014Atala,2014Huegel,2015Greschner_spont,2016Greschner,2020Buser}.
The other is the excitation energy $\Delta E$ [Fig.~\ref{fig_single}(c)], defined as $\Delta E=|\langle \psi(\tau)|\hat{H}|\psi(\tau)\rangle |-E_{g},$ where $E_g$ is the ground state energy for the final Hamiltonian. 
Both measures reflect the degree of adiabaticity observed in the fidelity.

The ultrafast adiabatic state preparation can be explained by the fact that the ground state does not depend on the flux for the choice $\theta_{\ell'\ell}^\parallel(\phi,\eta=0)$.
For the translationally invariant ladder, the single-particle Hamiltonian for quasimomentum $k$ reads $H(k)=h_0(k)+{\bm h}(k)\cdot {\bm{\sigma}}$ with $h_0(k)=-2J\cos(\phi/2)\cos(k+\eta)$, $h_x(k)=-J_\perp$, $h_y(k)=0$, $h_z(k)=-2J\sin(\phi/2)\sin(k+\eta)$, where the vector of Pauli matrices ${\bm \sigma}$ acts on the sublattice degree of freedom given by the upper and lower leg. 
The Bloch states $|\psi_\pm(k;\eta,\phi)\rangle$ of both bands $E_\pm(k)=h_0(k)\pm|{\bm h}(k)|$ are described by $k$ dependent vectors $\pm{\bm h}(k)/|{\bm h}(k)|$ on the Bloch sphere. In the MP the ground state lies at $k=-\eta/a$ with $h_z=0$. 
We define the overlap $\mathcal{O}_k(\phi)=|\langle\psi_-(k;0,0)|\psi_-(k;\eta,\phi)\rangle|^{2}$ to quantify the similarity between lowest-band eigenstates with and without magnetic flux $\phi$. 
Remarkably, in the case of $\eta=0$, the ground state wave function ($k=0$) does not depend on the magnetic flux $\phi$, as $h_z=h_y=0$ for all $\phi$ so that $\mathcal{O}_{k=0}(\phi)=1$ [Fig. \ref{fig_single}(d)].
For a system of $M$ rungs with periodic boundary condition, the quasimomentum $k$ takes discrete values given by integer multiples of $2\pi/M$.
As the spectrum is shifted by $\eta$, the squared overlap $\mathcal{O}\left(\pi/2\right)=\left|\left\langle \psi_{\pi/2}|\psi_0\right\rangle \right|^2$ between the initial and the target states drops suddenly from 1 to 0 when the shift $\eta$ becomes larger than $\pi/M$, as shown by the dashed line in Fig.~\ref{fig_single}(e).
Since $k$ is not a good quantum number anymore in the finite system with open boundary conditions, we observe a smooth decay of  $\mathcal{O}(\pi/2)$ as a function of $\eta$, starting from a value close to 1 for $\eta=0$ [$\mathcal{O}(\pi/2)=0.995$ for $M=24$ rungs]. This behaviour explains that the minimal ramping time $\tau^{*}$ required to reach $\mathcal{F}=0.9$ approaches zero when $\eta$ drops to zero.

\section{Comparison with counterdiabatic driving}

The idea of choosing an optimal vector potential for adiabatic state preparation can be related to the concept of counterdiabatic driving~\cite{2003Demirplak,2005Demirplak,2009Berry,2010Chen,2013Torrontegui,2019Odelin}. To be general, let $H_p$ be a Hamiltonian depending on a parameter $p$ and $|\psi_p\rangle$ the corresponding ground state. Starting from the ground state at $p=0$, we wish to rapidly prepare the ground state of the target Hamiltonian $H_{p=f}$. The idea of counterdiabatic driving is to consider a family of unitaries $U_p$ labelled by $p$, so that the evolved state exactly follows $|\psi_{p(t)}\rangle$ for a new Hamiltonian $H(t)=H_{p(t)}+i (d_t U_{p(t)})U_{p(t)}^\dagger$, where the second term corresponds to the so-called counterdiabatic driving that could be realized via some external forces~\cite{2003Demirplak,2005Demirplak}. Our approach, in turn, corresponds to directly working in the rotated frame of reference with instantaneous eigenstate $|\psi^{\prime}_p\rangle = U^{\dagger}_{p}|\psi_p\rangle$ governed by the Hamiltonian $H'_p=U^\dag_p H_p U_p$. For an ideal choice of $U_p$ (e.g.\ the optimal choice of Peierls phases with $\eta=0$ in this work), one can find a $p$-independent ground state $|\psi'_p\rangle=|\psi_0\rangle$, and thus it allows for the parameter ramp within arbitrarily short time. The advantage of our approach is that there will be no need for applying external terms to the system. Meanwhile, our protocol can be easily extended to the many-body system, as will be demonstrated in Sections \ref{sec_U} and \ref{sec_leaving}. %Namely, by making $|\psi'_p\rangle$ as close to $|\psi_0\rangle$ as possible, one can considerably speed up the adiabatic state preparation, as will be demonstrated in Figs.}

The optimal choice ($\eta=0$) of Peierls phases can also be understood intuitively by noting that the artificial electric fields generated during the ramp correspond to Faraday’s law of induction, as portrayed in Fig.~\ref{fig_single}(i).
In turn, for non-optimal choices with $\eta\neq 0$, additional electric fields are created as well during the ramp [causing the drift shown in Figs.~\ref{fig_single}(g,h)] that are not related to the time-dependence of the magnetic field via Faraday’s law of induction. These non-Faraday electric fields could be compensated by a time-dependent scalar potential.
This freedom of choosing $\eta$ is a consequence of the fact that the experimentalist directly engineers the artificial gauge potential (via Peierls phases) rather than the artificial magnetic field. The counterdiabatic driving terms required for rapid state preparation for the non-optimal choices of Peierls phases would simply correspond to time-dependent scalar potentials subtracting the non-Faraday forces generated by $\eta(t)\neq0$. 
Note, however, that the choice of $\eta=0$ and the absence of non-Faraday forces is not always optimal, as will be seen in Section \ref{sec_leaving} when discussing parameter ramps leaving the MP.

\begin{figure}
	\centering\includegraphics[width=0.98\linewidth]{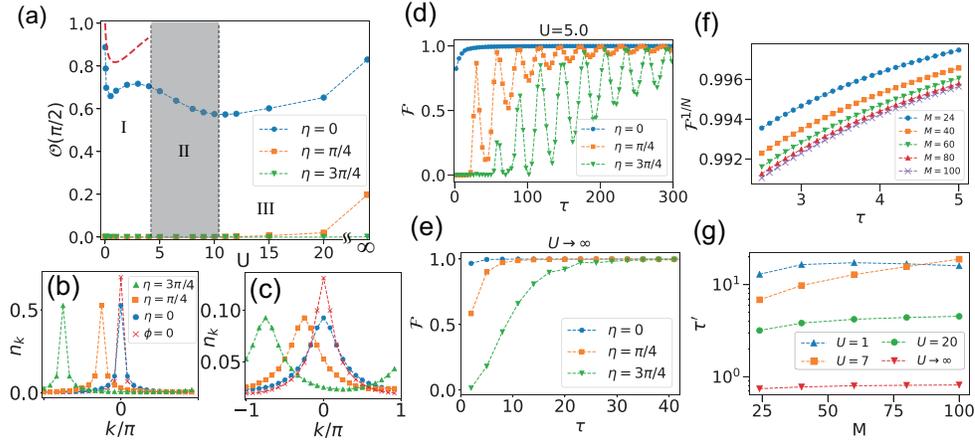}
	\caption{(a) Squared overlap of initial state $\left|\psi_0\right\rangle$ and target state  $\left|\psi_{\pi/2}\right\rangle$ as a function of $U$ for different values of $\eta$. The two vertical dashed lines $U_{c1}=4.2$ and $U_{c2}=10.4$ locate the BKT-transition points for $\phi=\pi/2$ and $\phi=0$ respectively. Quasimomentum distribution for (b) $U=1$ and (c) $U=10$. In the legend different $\eta$ refer to $\phi=\pi/2$. Fidelity as a function of total ramping time $\tau$ with interaction (d) $U/J=5$ and (e) hard-core limit.The simulations in (a-e) are performed in the system with the number of rungs $M=24$. (f) Single-particle fidelity $\mathcal{F}^{1/N}$ as a function of total ramping time $\tau$ for different numbers of rungs $M$ at $\eta=0$ and $U=20$. (g) Minimal ramping time $\tau^{\prime}$ required to reach $\mathcal{F}^{1/N}=0.995$ as a function of $M$ at $\eta=0$. For all cases we use $J_{\perp}=2$ at $1/2$ filling. (Dashed lines are guides to the eye.)} 
	\label{fig_ovlp}
\end{figure}

\section{Role of interactions}\label{sec_U}
Now we simulate the interacting system at filling $n=1/2$ per site by using the TeNPy library \cite{1992White,2005Schollwoeck,2013Kjaell,2018Hauschild} and a matrix product operator based time evolution method (tMPO)~\cite{2015Zaletel,2017Gohlke}. The ground state overlap $\mathcal{O}(\pi/2)$ as a function of interaction strength $U$ is plotted in Fig.~\ref{fig_ovlp}(a).
In the case of $\eta=0$, the overlap $\mathcal{O}(\pi/2)$ exhibits non-monotonous behavior, reflecting a complex competition between many-body interactions and artificial magnetic flux.
While the system features a Meissner-like superfluid ground state for weak interactions~\cite{supp}, (in the thermodynamic limit) it undergoes a Berezinskii-Kosterlitz-Thouless (BKT) transition to a Mott-insulator state with single particles localised on the rungs as $U$ is increased~\cite{1973Kosterlitz,2012Dhar,2016Greschner}.
The critical parameter is found to be $U_{c1}\approx4.2$ for $\phi=\pi/2$ and $U_{c2}\approx10.4$ for $\phi=0$ \cite{supp}, which determines three regions (I: $U<U_{c1}$, II: $U_{c1}<U<U_{c2}$, and III: $U_{c2}<U$) shown in Fig.~\ref{fig_ovlp}(a), where we plot the overlap $\mathcal{O}(\pi/2)$ (blue dots connected by dashed line). 
In the weakly interacting region I, the overlap first decreases rapidly, before it slightly increases again. This behaviour is qualitatively reproduced by Bogoliubov theory (red dashed line)~\cite{supp}. It can be related to the fact that the interaction-induced population of finite momentum modes initially happens much faster in the presence of magnetic flux (giving rise to an enlarged effective mass). However, for even stronger interactions the resulting momentum mismatch becomes smaller again~\cite{supp}. 
For $U_{c1}<U<U_{c2}$, while the ground state with zero flux remains superfluid, the ground state with flux $\phi=\pi/2$ already becomes a Mott insulator~\cite{supp}, and therefore the overlap decreases once more. After $U>U_{c2}$, the fact that both ground states present Mott-insulating phase gives rise to an increase again.
Despite this non-monotonous behavior, $\mathcal{O}(\pi/2)$ takes comparably large values for $\eta=0$.
This leads to rather short adiabatic preparation times also in the strongly interacting regime.
In Figs.~\ref{fig_ovlp}(d) and (e), we plot the fidelity $\mathcal{F}$ versus the ramping time for $U=5$ and the hard-core limit $U\rightarrow\infty$, respectively. Remarkably, for hard-core bosons (and $\eta=0$), we find fidelities close to one already for very short ramping times on the order of 1 (in units of the tunnelling time). The fact that such rapid state preparation found for the strongly interacting system cannot be explained by the single-particle analysis presented in the previous sections. This short ramping time may be related to the fact that the larger overlap of quasimomentum distribution occurs for stronger interaction, as indicated in Figs~\ref{fig_ovlp}(b) and (c).

In  Figs.~\ref{fig_ovlp}(f) and (g), we investigate the finite size effect for the optimal choice of $\eta=0$. The fact that a many-body fidelity drops with the system size can be attributed to two effects. On the one hand, a finite-size gap (as present in the superfluid regime) separating the ground state from the first excited state decreases with ladder length $M$, leading to a reduction of adiabaticity. On the other hand, a decrease of the many-body fidelity with system size is expected already from the very fact that (at least for product states) the $N$-particle fidelity is given by the $N$th power of the single-particle fidelity. In order to compensate for the latter effect, when comparing results for different system sizes, we use the single-particle fidelity $\mathcal{F}^{1/N}$. In Fig. 3(f), we plot $\mathcal{F}^{1/N}$ versus the ramping time $\tau$ for $\eta=0$ and $U=20$ for different ladder length $M$. In Fig. 3(g), we extract the ramping time $\tau^{\prime}$ above which a fidelity $\mathcal{F}^{1/N}\geq 0.995 $ is achieved and plot it versus $M$ for various interaction strengths $U$. Remarkably, we find that the ramping time increases very slowly both in the superfluid and the Mott-insulating regions (i.e.\ I and III). A noticeable increase is only visible in regime II, where the Mott transition occurs during the ramp.

For finite values of $\eta$, taking $\eta=\pi/4$, $3\pi/4$ as examples shown in Fig.~\ref{fig_ovlp}(a), $\mathcal{O}(\pi/2)$ takes small values until deep in the Mott regime, where the correlations between individual rungs are suppressed by interactions for both $\phi=0$ and $\phi=\pi/2$. 
This can also be understood from the quasi-momentum distribution defined by $n_{k}=\frac{1}{M}\sum_{n=0,1}\sum_{m,m^{\prime}}e^{ik(m-m^{\prime})}\langle \hat{a}_{m^{\prime},n}^{\dagger}\hat{a}_{m,n}\rangle$.
From Fig.~\ref{fig_ovlp}(b,c) we can see that the distribution is centered around $k=0$ for the initial state ($\phi=0$), and at $k=-\eta$ for the target state ($\phi=\pi/2$). 
Although the shift of quasi-momentum (for $\eta\neq 0$) causes difficulties in state preparations, the increase of interaction broadens the quasimomentum distributions, which results in gradually increasing overlap and a shorter adiabatic ramping time as indicated in Fig.~\ref{fig_ovlp}(d,e). 

\section{Leaving the Meissner regime}\label{sec_leaving}
\begin{figure}
	\centering\includegraphics[width=0.75\linewidth]{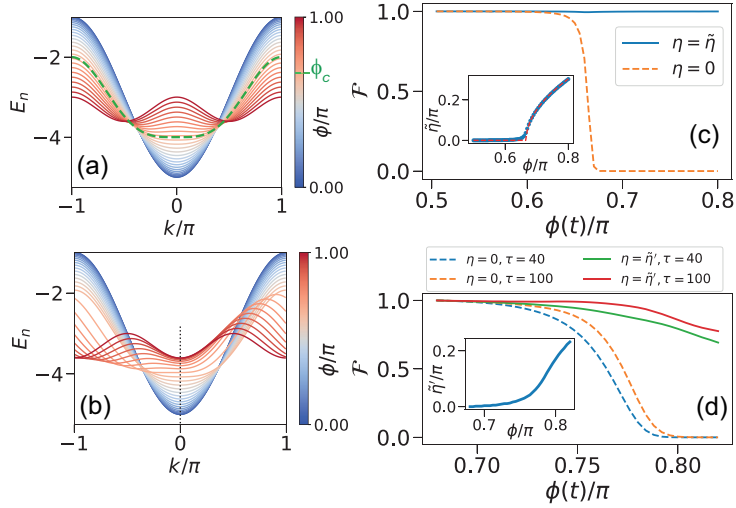}
	\caption{(a) Spectrum of the non-interacting ladder with $\theta^{\parallel}_{\ell'\ell}(\eta=0)$ at $J_{\perp}=3$. (b) Spectrum of the non-interacting ladder with $\theta^{\parallel}_{\ell'\ell}(\eta=\tilde{\eta})$, where $\tilde{\eta}$ shifts the spectrum so that the right minimum is always located at $k=0$. (c) The fidelity as a function of time-dependent flux $\phi(t)$ at $U=0$. The flux is ramped from $\phi_i=0.5\pi$ to $\phi_f=0.8\pi$ within a ramping time $\tau=30$ in a ladder with $M=50$ rungs at $J_{\perp}=3$. The inset depicts $\tilde{\eta}$ as a function of $\phi$, where the blue dots come from maximizing the ground state overlap $\mathcal{O}(\phi)$ and the red dashed line is the analytical results. (d) The total fidelity as a function of $\phi(t)$ at $U=2.0,J_{\perp}=3$ with particle number of $N=40$ in the ladder with $M=25$ rungs.  The flux is ramped from $\phi^{\prime}_i=0.68\pi$ to $\phi^{\prime}_f=0.82\pi$ within time $\tau=40$ and $100$. The inset depicts $\tilde{\eta}^{\prime}$ as a function of $\phi$ which maximizes the ground state overlap.}
	\label{fig_cross}
\end{figure}

So far, we considered parameter ramps within the MP. Increasing $\phi$ further gives rise to various phases~\cite{2015Piraud,2015Greschner_spont,2016Greschner,2020Buser},
including the biased ladder phase (BLP) in the weakly and intermediately interacting regime~\cite{2014Wei,2015Uchino,2016Uchino,2015Greschner_spont,2016Greschner,2020Buser}, which is characterized by vanishing rung currents and the spontaneous $\mathbb{Z}_2$ reflection symmetry breaking in the form of a density imbalance between both legs. In the following, we show that starting from the MP, the BLP can be efficiently prepared by choosing proper Peierls phase patterns (determined by $\eta$).
Let us start with the non-interaction limit, where beyond a critical flux $\phi_c$, the system enters the vortex phase and the dispersion relation develops two degenerate minima. 
Since each minimum predominantly corresponds to the occupation of one of the legs, the degeneracy can be lifted by introducing a small bias potential ($0.01J$) between both legs, so that the ground state resembles that of the BLP.
Despite the fact that the small bias softens the sharp transition at $\phi_c\approx0.667\pi$ into a narrow crossover, we observe a sudden drop of the fidelity at $\phi_c$ when linearly ramping up the Peierls phases with $\eta=0$ [Fig.~\ref{fig_cross}(c)].
Here the dashed line represents the fidelity between the evolved state and the instantaneous eigenstate.
As a remedy, one can vary $\eta$ during the ramp in such a fashion that the overlap $\mathcal{O}(\phi)$ remains maximal during the ramp. 
(For an infinitely large system without bias, this can be achieved by choosing $\tilde{\eta}(t)=\arccos\sqrt{J_{\perp}^{2}/4\cot^{2}(\phi/2)+\cos^{2}(\phi/2)} $ for $ \phi>\phi_{c}$, so that the right minimum of the dispersion relation always remains at $k=0$ [Fig.~\ref{fig_cross}(b)].)
In this case, the evolved state successfully follows the instantaneous eigenstate even after the critical point, as indicated by the horizontal blue line in Fig.~\ref{fig_cross}(c).
Thus, different from the previously discussed case, now the optimal choice of Peierls phases does not correspond to the situation where all the non-Faraday forces were absent during the ramp. Instead, the forces induced by $\eta\neq0$ are actively employed for state preparation, as they induce shifts in quasimomentum that keep the system state at the minimum of the dispersion relation.

The scheme can also be applied to the interacting system. For instance, the transition to the BLP occurs at critical flux $\phi^{\prime}_{c}\approx0.8\pi$ for a 0.8-filling ladder at $U=2.0,J_{\perp}=3$~\cite{2016Greschner}. As shown by the dashed lines in Fig.~\ref{fig_cross}(d), using $\eta=0$ leads to an essentially vanishing fidelity after the critical point, as the BLP has imbalanced distribution between positive and negative quasimomenta due to the broken reflection symmetry~\cite{2014Wei,2020Buser}. To compensate the quasimomentum differences between initial and final states during the ramp, the protocol $\tilde{\eta}^{\prime}(t)$ [shown in the inset of Fig.~\ref{fig_cross}(d)] can be determined by maximizing the ground state overlap, and the corresponding $\mathcal{F}$ assumes rather large values as shown by solid lines in Fig.~\ref{fig_cross}(d). Note that the finite value $\mathcal{F}=0.78$ found for $\tau=100$ indicates a near unity single-particle fidelity ($0.78\approx0.994^{N}$) for the system with number of particle $N=40$ considered here. Higher fidelities can be achieved for longer ramping times.

\section{Conclusion and Outlook}
We have proposed to design the time-dependent artificial vector potentials in the form of Peierls phases for rapid adiabatic state preparation in optical lattice systems. Our approach is based on the fact that in such systems the experimentalist directly controls the vector potential rather than magnetic fields. We demonstrated that for a ladder with flux, this approach allows for an almost immediate state preparation for non-interacting bosons. Remarkably, we find very short ramping times also for strongly interacting bosons. 

While the abrupt adiabatic preparation in the ladder is an extreme example, it highlights that tunning Peierls phases can be a very powerful tool for state preparation.
Specifically, choosing optimal gauge potentials to maximize the overlap between the instantaneous eigenstate and the initial state helps to reduce adiabatic ramping time.
It is an interesting open question for future research in how far this approach can be used for the preparation of strongly correlated states of matter, such as fractional Chern insulators.

\section*{Acknowledgements}
We thank Monika Aidelsburger, Maximilian Buser, Andrew Hayward, Julian L\'{e}onard, Fabian Heidrich-Meisner and Frank Pollmann for discussions. The research was funded by the Deutsche Forschungsgemeinschaft (DFG) via the Research Unit FOR 2414 under Project No. 277974659. Xiao-Yu Dong was supported by the U.S. Department of Energy, Office of Science, Advanced Scientific Computing Research and Basic Energy Sciences, Materials Sciences and Engineering Division, Scientific Discovery through Advanced Computing (SciDAC) program under the grant number DE-AC02- 76SF00515. F.~N.~\"U.~acknowledges support from the Royal Society under the Newton International Fellowship.

\section{Appendix}

\subsection{Dynamics during the ramp}
\begin{figure}
	\centering\includegraphics[width=0.75\linewidth]{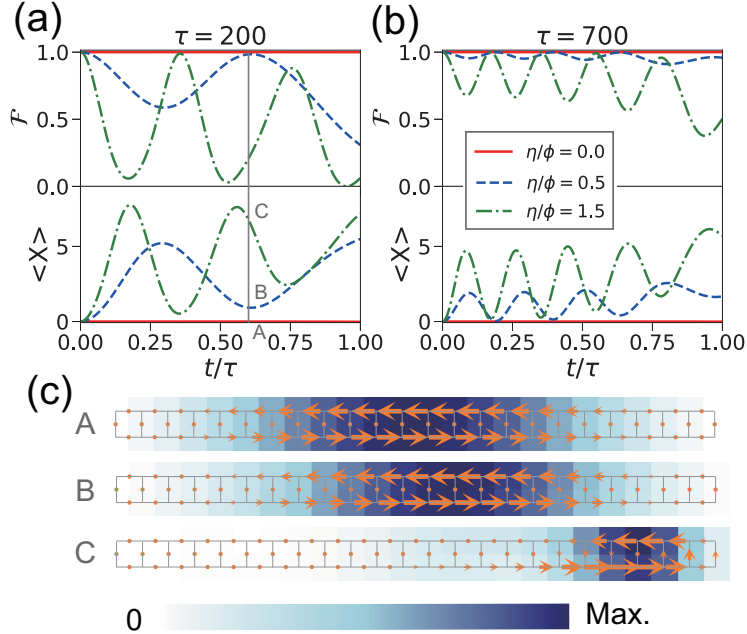}
	\caption{Fidelity $\mathcal{F}$ (upper panel) and center of mass $\langle X \rangle$ (lower panel) as a function of time within (a) $\tau=200$ and (b) $\tau=700$. The origin $\langle X \rangle=0$ is defined to lie at the middle of the ladder. (c) Spatial density and probability current distributions at $t/\tau=0.6$ with $\tau=200$. A-C correspond to $\eta/\phi=0,0.5,1.5$ respectively. It shows that the closer of center-of-mass to the middle of ladder, the larger fidelity is obtained. Other parameters are chosen as $U=0,~J_{\perp}=2,~\phi(t)=(\pi/2)t/\tau$ and $M=24$.} 
	\label{fig_evo}
\end{figure}

Each point in Fig.~2(a-c) in main text corresponds to the result at the end of a parameter ramp. 
To interpret the oscillation behavior, we plot the fidelity $\mathcal{F}=\left|\left\langle \psi_{\phi}|\psi(t)\right\rangle \right|^2$ and center of mass $\langle X \rangle$ during a single ramping process in Fig.~\ref{fig_evo}.
It shows that while the center of mass gets closer to the middle of the ladder, the fidelity always has a large value.
Thus the oscillation of $\mathcal{F}$ is related to the Bloch oscillations of the atomic cloud, which are triggered by the non-Faraday synthetic electric fields that are generated during the ramp for non-zero $\eta$.
% For finite value of $\eta$, although the fidelity can be close to $1$ at a comparably short time $\tau$, it goes down for a longer time. This indicates a not well prepared state in spite of a large value of fidelity. In the case of $\eta=0$, however, the absence of any oscillation during the ramp demonstrates a successful adiabatic preparation.

\subsection{Meissner-like phases}
\begin{figure}
	\centering\includegraphics[width=0.7\linewidth]{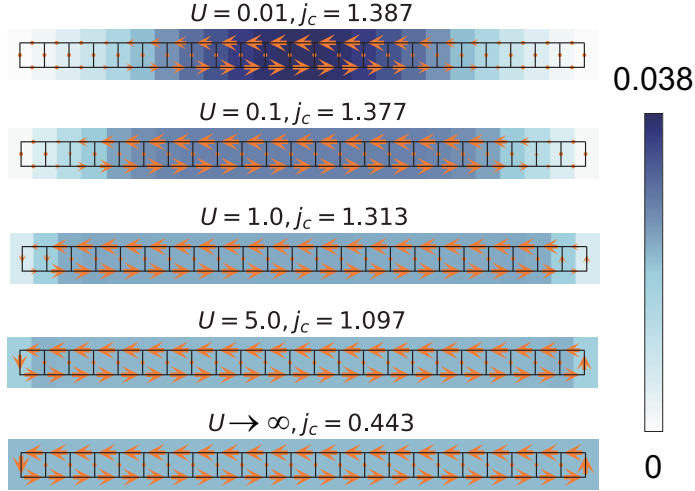}
	\caption{Probability current patterns for different $U$ at $J_{\perp}=2, \phi=\pi/2, n=1/2, M=24$. The arrow size is proportional to the expectations values of the local currents. }
	\label{fig_currents}
\end{figure}

The ground state chiral current can be used to characterize different phases in a ladder system, like the Meissner or vortex phase.
Based on the continuity relation, the local current operators on legs and rungs are respectively defined as~\cite{1991Silva,2014Huegel,2016Greschner,2019Buser},
\begin{eqnarray}
& \hat{j}_{m,n}^{\parallel}= iJ\left(e^{-i\left(\phi\left(1/2-n\right)-\eta\right)}\hat{a}_{m,n}^{\dagger}\hat{a}_{m+1,n}-h.c.\right),
\\
& \hat{j}_{m}^{\perp}= iJ_{\perp}\left(\hat{a}_{m,0}^{\dagger}\hat{a}_{m,1}-h.c.\right),
\end{eqnarray}
which gives the global chiral current $j_{c}=\frac{1}{M}\sum_{m=0}^{M-2}\left\langle \hat{j}_{m,0}^{\parallel}-\hat{j}_{m,1}^{\parallel}\right\rangle $.
%The vanishing rung currents in Fig.~\ref{fig_scheme}(f) characterize a Meissner state in a finite system with open boundary condition.
At small fluxes, probability currents exist only along the legs and behave like screening currents, thus the low-flux phase is identified as a Meissner phase, in analogy to that in a type-II superconductor. For large values of the flux, the system enters into a vortex phase, where finite rung currents emerge and form vortex structures. From Fig.~\ref{fig_currents} we can see that for $J_{\perp}/J=2,\phi=\pi/2$, the system with finite size assumes a Meissner-like phase for various values of $U$.

\subsection{Bogoliubov Theory}
The Hamiltonian can be written as
\begin{eqnarray}
\hat{H}=&\hat{H}_{S}+\hat{H}_{I},
\\ 
\hat{H}_{S}=&-J{\displaystyle \sum_{r}}\left(e^{i\theta_{1}}\hat{a}_{1,r+1}^{\dagger}\hat{a}_{1,r}+e^{i\theta_{2}}\hat{a}_{2,r+1}^{\dagger}\hat{a}_{2,r}+h.c.\right)
\nonumber \\ &
-J_{\perp}{\displaystyle \sum_{r}}\left(\hat{a}_{2,r}^{\dagger}\hat{a}_{1,r}+\hat{a}_{1,r}^{\dagger}\hat{a}_{2,r}\right),
\\
\hat{H}_{I}=&\frac{U}{2}{\displaystyle \sum_{r}}\left(\hat{a}_{1,r}^{\dagger}\hat{a}_{1,r}^{\dagger}\hat{a}_{1,r}\hat{a}_{1,r}+\hat{a}_{2,r}^{\dagger}\hat{a}_{2,r}^{\dagger}\hat{a}_{2,r}\hat{a}_{2,r}\right).
\label{H_al}
\end{eqnarray}
Here $\hat{a}_{1,r}^{\dagger}$ ($\hat{a}_{1,r}$) and  $\hat{a}_{2,r}^{\dagger}$ ($\hat{a}_{2,r}$) are the creation (annihilation) operators on the rung $r$ in the lower and upper leg respectively, $J$ denotes the amplitude of nearest-neighbor tunneling along the legs, with $\theta_{1,2}=-\eta\pm\phi/2$ being the corresponding Peierls phases, so that the flux in each plaquette is $\phi$ and we consider $\phi=\pi/2$ here. %$J_{\perp}$ is the hopping amplitude along the rungs, $h.c.$ means the Hermitian conjugate, $U$ is the on-site repulsive interaction energy.

For a two-leg ladder with $M$ rungs, under periodic boundary conditions along the legs, the quasimomentum takes discrete value $k=\frac{2\pi}{Ma}m$ with $m=0,\pm1,\pm2,\cdots,\pm M/2$ and $a$ being the lattice constant. By performing the Fourier transformation
\begin{equation}
\hat{a}_{l,r}=\frac{1}{\sqrt{M}}\sum_{k}e^{ikar}\hat{a}_{l,k},~ l=1,2
\end{equation}
the above Hamiltonians can be expressed in quasi-momentum representation as
\begin{align}
\hat{H}_{S}= & {\displaystyle \sum_{k}}\left(\epsilon_{1,k}\hat{a}_{1,k}^{\dagger}\hat{a}_{1,k}+\epsilon_{2,k}\hat{a}_{2,k}^{\dagger}\hat{a}_{2,k}\right)
-J_{\perp}{\displaystyle \sum_{k}}\left(\hat{a}_{2,k}^{\dagger}\hat{a}_{1,k}+\hat{a}_{1,k}^{\dagger}\hat{a}_{2,k}\right),
\label{HS_k} \\ 
\hat{H}_{I}= &\frac{U}{2M}{\displaystyle \sum_{\{k_i\}}}\left(\hat{a}_{1,k_1}^{\dagger}\hat{a}_{1,k_2}^{\dagger}\hat{a}_{1,k_3}\hat{a}_{1,k_4}+\hat{a}_{2,k_1}^{\dagger}\hat{a}_{2,k_2}^{\dagger}\hat{a}_{2,k_3}\hat{a}_{2,k_4}\right)
\times \tilde{\delta}_{k_1+k_2,k_3+k_4},
\label{HI_k}
\end{align}
with
\begin{eqnarray}
\epsilon_{1,k}=-2J\cos\left(ka+\eta-\phi/2\right),
\label{epsilon_1}
\\
\epsilon_{2,k}=-2J\cos\left(ka+\eta+\phi/2\right),
\label{epsilon_2}
\end{eqnarray}
and periodic Kronecker symbol $\tilde{\delta}_{k,q}$ vanishing unless $k=q$ modulo reciprocal lattice constants $2\pi/a$.

\subsubsection{Diagonal basis}
The single-particle Hamiltonian (\ref{HS_k}) can be diagonalized by choosing a different basis, i.e.
\begin{eqnarray}
\left(\begin{array}{c}
\hat{a}_{1,k}\\
\hat{a}_{2,k}
\end{array}\right)=\left(\begin{array}{cc}
u_{k} & -v_{k}\\
v_{k} & u_{k}
\end{array}\right)\left(\begin{array}{c}
\hat{b}_{1,k}\\
\hat{b}_{2,k}
\end{array}\right).
\label{a_b}
\end{eqnarray}
The canonical commutation $\left[\hat{a}_{k},\hat{a}_{k^{\prime}}^{\dagger}\right]=\delta_{k,k^{\prime}}$ requires that 
\begin{eqnarray}
u_{k}^{2}+v_{k}^{2}=1.
\end{eqnarray}
%Note that here it is NOT a Bogoliubov transformation. 
Substituting Eq.~(\ref{a_b}) to Eq.(\ref{HS_k}), and imposing all the off-diagonal terms to vanish, the single particle Hamiltonian is diagonalized as
\begin{eqnarray}
\hat{H}_{S}=E_{+}\hat{b}_{1,k}^{\dagger}\hat{b}_{1,k}+E_{-}\hat{b}_{2,k}^{\dagger}\hat{b}_{2,k},
\end{eqnarray}
with
\begin{eqnarray}
& E_{+}=\frac{1}{2}\left(\epsilon_{1,k}+\epsilon_{2,k}+\sqrt{4J_{\perp}^{2}+\left(\epsilon_{1,k}-\epsilon_{2,k}\right)^{2}}\right),\\
& E_{-}=\frac{1}{2}\left(\epsilon_{1,k}+\epsilon_{2,k}-\sqrt{4J_{\perp}^{2}+\left(\epsilon_{1,k}-\epsilon_{2,k}\right)^{2}}\right),\\
& u_{k}^{2}=\frac{1}{2}\left(1-\frac{\epsilon_{2,k}-\epsilon_{1,k}}{\sqrt{4J_{\perp}^{2}+\left(\epsilon_{1,k}-\epsilon_{2,k}\right)^{2}}}\right).
\label{uk2vk2}
\end{eqnarray}

\subsubsection{Truncation to the lowest band}
The terms related to $\hat{b}_{2,k}$ ($\hat{b}_{1,k}$) correspond to the lower (upper) band. Since the system possesses a large band gap for the parameters used ($J_\perp=2J$), for weak interaction we are allowed to truncate our Hamiltonian to the lowest band. To do this we substitute Eq.~(\ref{a_b}) into the Hamiltonian and neglect the $\hat{b}_{1,k}$ terms.
In this case, the full Hamiltonian is truncated to the lowest band \cite{2014Wei},
\begin{equation}
\hat{H}= {\displaystyle \sum_{k}}E_{-}\left(k\right)\hat{b}_{k}^{\dagger}\hat{b}_{k}
+\frac{U}{2M}{\displaystyle \sum_{\{k_i\}}}\Gamma_{k_1,k_2,k_3,k_4}\hat{b}_{k_1}^{\dagger}\hat{b}_{k_2}^{\dagger}\hat{b}_{k_3}\hat{b}_{k_4}\tilde{\delta}_{k_1+k_2,k_3+k_4},
\label{H_trun}
\end{equation}
where $\hat{b}_k \equiv \hat{b}_{2,k}$ and we have defined $\Gamma_{k_1,k_2,k_3,k_4}=v_{k_1}v_{k_2}v_{k_3}v_{k_4}+u_{k_1}u_{k_2}u_{k_3}u_{k_4}$.

\subsubsection{Bogoliubov approximation}
For weak interactions and at low temperature, the number $N_0$ of particles occupying the single-particle ground state with quasi momentum $k_0$ remains of the order of total particle number $N$ in a system of finite extent. Thus one can make the approximation
\begin{equation}
\hat{N}_{0}=\hat{b}_{k_0}^{\dagger}\hat{b}_{k_0}\simeq \hat{N}_{0}+1=\hat{b}_{k_0}\hat{b}_{k_0}^{\dagger},
\end{equation}
which leads to
\begin{eqnarray}
& \hat{b}_{k_0}\simeq\hat{b}_{k_0}^{\dagger}=\sqrt{N_{0}},
\\ &
\hat{b}_{k}=\sqrt{N_{0}}\delta_{k,k_{0}}+\hat{b}_{k}\left(1-\delta_{k,k_{0}}\right).
\label{Bogo_presc}
\end{eqnarray}
Keeping all the terms up to second order in $\hat{b}_{k\neq k_0}$, the Hamiltonian (\ref{H_trun}) becomes
\begin{eqnarray}
\hat{H}= &
E_{-}\left(k_0\right)N_{0}+\frac{U}{2M}\Gamma_{0}{UN_{0}^2}+{\displaystyle \sum_{k\neq0}}E_{-}\left(k+k_{0}\right)\hat{b}_{k}^{\dagger}\hat{b}_{k}
\nonumber \\ &
+\frac{UN_{0}}{2M}{\displaystyle \sum_{k\neq0}}\left[\Gamma_{1}\left(\hat{b}_{k}\hat{b}_{-k}+\hat{b}_{k}^{\dagger}\hat{b}_{-k}^{\dagger}\right)+4\Gamma_{2}\hat{b}_{k}^{\dagger}\hat{b}_{k}\right],
\label{H_N0}
\end{eqnarray}
with the coefficients 
\begin{eqnarray}
& \Gamma_{0}= v_{k_{0}}^{4}+u_{k_{0}}^{4}=1/2,\\
& \Gamma_{1}= \left(v_{k+k_{0}}v_{k_{0}-k}+u_{k+k_{0}}u_{k_{0}-k}\right)/2,\\
& \Gamma_{2}= v_{k_{0}}^{2}v_{k+k_{0}}^{2}+u_{k_{0}}^{2}u_{k+k_{0}}^{2}=1/2,
\label{Gamma_012}
\end{eqnarray}
where we have used $v_{k_{0}}^{2}=1/2=u_{k_{0}}^{2}$ according to Eqs.~(\ref{epsilon_1}), (\ref{epsilon_2}) and (\ref{uk2vk2}).

Substituting $N_{0}=N-\sum_{k\neq0}\hat{b}_{k}^{\dagger}\hat{b}_{k}$ and keeping the terms up to second order in $\hat{b}_{k}$, we arrive at
\begin{eqnarray}
\hat{H}= & E_{0}-{\displaystyle \sum_{k>0}}C_{-k}+{\displaystyle \sum_{k>0}}\hat{H}_{k}
\nonumber \\
\hat{H}_{k}= &\left(\begin{array}{cc}
\hat{b}_{k}^{\dagger} & \hat{b}_{-k}\end{array}\right)\left(\begin{array}{cc}
C_{k} & 2D_{k}\\
2D_{k} & C_{-k}
\end{array}\right)\left(\begin{array}{c}
\hat{b}_{k}\\
\hat{b}_{-k}^{\dagger}
\end{array}\right),
\label{Hk_positive}
\end{eqnarray}
with 
\begin{eqnarray}
& E_{0}=\left(E_{-}\left(k_0\right)+Un\Gamma_{0}\right)N, \\
& C_{k}=E_{-}\left(k+k_{0}\right)-E_{-}\left(k_0\right)+Un\left(4\Gamma_{2}-2\Gamma_{0}\right),\\
& D_{k}=Un\Gamma_{1}=D_{-k}\equiv D.
\label{E0CD}
\end{eqnarray}
Here we have introduced the total particle number per site $n=\frac{N}{2M}$, and the additional term $-\sum_{k>0}C_{-k}$ comes from the commutation relation $\hat{b}_{-k}^{\dagger}\hat{b}_{-k}=\hat{b}_{-k}\hat{b}_{-k}^{\dagger}-1$. 

\subsubsection{Diagonalization}
To diagonalize the Hamiltonian (\ref{Hk_positive}), we perform the Bogoliubov transformation
\begin{equation}
\left(\begin{array}{c}
\hat{b}_{k}\\
\hat{b}_{-k}^{\dagger}
\end{array}\right)=\left(\begin{array}{cc}
\mu & \nu\\
\nu & \mu
\end{array}\right)\left(\begin{array}{c}
\hat{\rho}_{k}\\
\hat{\rho}_{-k}^{\dagger}
\end{array}\right),
\label{transf}
\end{equation}
with quasiparticle annihilation (creation) operators $\hat{\rho}_k$ ($\hat{\rho}^{\dagger}_k$).
Requiring bosonic commutation relations for the quasiparticle operators, we have
\begin{eqnarray}
\mu^{2}-\nu^{2}=1.
\label{uk_vk_bar}
\end{eqnarray}

To get the expressions for $\mu,\nu$, we plug Eq.~(\ref{transf}) into Eq.~(\ref{Hk_positive}) and impose that 
\begin{eqnarray}
\hat{H}_{k}=\left(\begin{array}{cc}
\hat{\rho}_{k}^{\dagger} & \hat{\rho}_{-k}\end{array}\right)\left(\begin{array}{cc}
\gamma_{1} & 0\\
0 & \gamma_{2}
\end{array}\right)\left(\begin{array}{c}
\hat{\rho}_{k}\\
\hat{\rho}_{-k}^{\dagger}
\end{array}\right). 
\label{Hk_diag}
\end{eqnarray}
Thus we have 
\begin{eqnarray}
\left(\begin{array}{cc}
\gamma_{1} & 0\\
0 & \gamma_{2}
\end{array}\right)=\left(\begin{array}{cc}
\mu & \nu\\
\nu & \mu
\end{array}\right)\left(\begin{array}{cc}
C_{k} & 2D_{k}\\
2D_{k} & C_{-k}
\end{array}\right)\left(\begin{array}{cc}
\mu & \nu\\
\nu & \mu
\end{array}\right)
\end{eqnarray}
which leads to the solutions:
\begin{eqnarray}
& \gamma_{1}=\frac{1}{2}\left(C_{k}-C_{-k}+\sqrt{\left(C_{-k}+C_{k}\right)^{2}-16D^{2}}\right),\\
& \gamma_{2}=\frac{1}{2}\left(-C_{k}+C_{-k}+\sqrt{\left(C_{-k}+C_{k}\right)^{2}-16D^{2}}\right),\\
& \mu^{2}=\frac{1}{2}\left(1+\frac{C_{-k}+C_{k}}{\sqrt{\left(C_{-k}+C_{k}\right)^{2}-16D^{2}}}\right).
\label{gamma12}
\end{eqnarray}

\subsubsection{Bogoliubov ground state}
In the following, we follow Ref.~\cite{2010Ueda} and construct the Bogoliubov ground state $\left|\Psi_{0}^{B}\right\rangle $, which is defined as the state with no quasi-particle, i.e.
\begin{eqnarray}
\hat{\rho}_{k}\left|\Psi_{0}^{B}\right\rangle =0,~\forall k\ne k_0
.
\label{psi_bogo_def}
\end{eqnarray}
As the Bogoliubov transformation (\ref{transf}) connects the states with $k$ and $-k$, the Bogoliubov ground state can be expressed as the states where $n_k$ particles are present in $k$ states and $n_{-k}$ particles are in the $-k$ states \cite{2010Ueda}, i.e.
\begin{eqnarray}
\left|\Psi_{0}^{B}\right\rangle =\prod_{k}\sum_{n,n_{-k}}C_{n_k,n_{-k}}^{k}\frac{(\hat{b}^{\dagger}_k)^{n_k}}{\sqrt{n_k !}}
\frac{(\hat{b}^{\dagger}_{-k})^{n_{-k}}}{\sqrt{n_{-k} !}}
|0\rangle,
\label{psi_Bogo}
\end{eqnarray}
where $|0\rangle$ denotes the vacuum state.~Substituting Eq.~(\ref{psi_Bogo}) into Eq.~(\ref{psi_bogo_def}) and using the expression of $\hat{\rho}_{k}=\mu\hat{b}_{k}-\nu\hat{b}_{-k}^{\dagger}$ according to Eq.~(\ref{transf}), we have
\begin{equation}
\prod_{k}\sum_{n_{k},n_{-k}=0}^{\infty}\left(C_{n_{k}+1,n_{-k}}^{k}\mu\sqrt{n_{k}+1}\right.
+\left.C_{n_{k},n_{-k}-1}^{k}-\nu\sqrt{n_{-k}}\left|n_{k},n_{-k}\right\rangle \right)=0,
\end{equation}
where we define $C_{n_k,-1}^{k}=0$. Since the basis $\left\{ \left|n_k,n_{-k}\right\rangle \right\} $ are orthogonal, we get
\begin{equation}
\sqrt{n_k+1}C_{n_k+1,n_{-k}}^{k}+\alpha_{k}\sqrt{n_{-k}}C_{n_k,n_{-k}-1}^{k}=0
\label{C_nm}
\end{equation}
with $\alpha_{k}=-\nu/\mu$ for short.

By setting $n_{-k}=0$ in the above equation Eq.~(\ref{C_nm}), we have $C_{n_k+1,0}^{k}=0~\left(n_k\geq0\right)$.
The similar procedure for $\hat{b}_{1,-k}\left|\Psi_{0}^{B}\right\rangle =0$ gives us $C_{0,n_{-k}+1}^{k}=0~\left(n_{-k}\geq0\right)$. Based on these observations, it turns out that all the `off-diagonal' components vanish, i.e. $C_{n_k+1,n_{-k}}^{k}=0 ~(n_{-k} \neq n_k+1)$. In the case of $n_{-k}=n_k+1$, Eq.~(\ref{C_nm}) gives us the following expression of the diagonal terms
\begin{eqnarray}
C_{n_k,n_k}^{k}=\left(-\alpha_{k}\right)^{n_k}C_{0,0}^{k},
\end{eqnarray}
where $C_{0,0}^{k}$ is determined from the normalization of the wave-function.
Therefore, the Bogoliubov ground state is a state where pairs of particles with wave vector $k$ and $-k$ are excited. 

We denote $\left|n_{1,}n_{2},\cdots\right\rangle $ as a state with $n$ pairs of particles with non-zero quasi-momentum $k$ and $-k$, and $\left|\psi_{0}\right\rangle $ as the state with $k=0$. In this case the Bogoliubov ground state takes the following form
\begin{eqnarray}
\left|\Psi_{0}^{B}\right\rangle =Z\sum_{n_{1},n_2}\left[\left(-\alpha_{k_{1}}\right)^{n_{1}}\left(-\alpha_{k_{2}}\right)^{n_{2}}\cdots\right]\left|n_{1},n_{2},\cdots\right\rangle \left|\psi_{0}\right\rangle,
\label{bogo_ground_stat}
\end{eqnarray}
where $Z={\displaystyle \prod_{k>0}}\sqrt{1-\alpha_{k}^{2}}$ is the normalization factor.

The state $\left|\psi_{0}\right\rangle $ for $k=0$ is a coherent state $\hat{b}_{0}\left|\psi_{0}\right\rangle=\psi_{0}\left|\psi_{0}\right\rangle $ and reads
\begin{eqnarray}
\left|\psi_{0}\right\rangle
=Z_{0}\sum_{n_{0}}\frac{\psi_{0}^{n_{0}}}{\sqrt{n_{0}!}}\left|n_{0}\right\rangle ,
\label{coherentstate}
\end{eqnarray}
where we have defined the vacuum state $\left|\text{vac}\right\rangle$ for the real particles operators $\hat{b}_{k}$, i.e. $\hat{b}_{k}\left|\text{vac}\right\rangle=0$. The normalization factor is $Z_{0}=\exp\left(-\left|\psi_{0}\right|^{2}/2\right)$.

According to Eq.~(\ref{bogo_ground_stat}) we have the overlap of two ground states
\begin{eqnarray}
\mathcal{O} =\left\langle \Psi_{0}^{\prime B}\right.\left|\Psi_{0}^{B}\right\rangle =ZZ^{\prime}\left\langle \psi_{0}^{\prime}\right.\left|\psi_{0}\right\rangle {\displaystyle \prod_{k>0}}\frac{1}{1-\alpha_{k}^{\prime}\alpha_{k}}.
\label{overlap}
\end{eqnarray}
The overlap of coherent states $\mathcal{O}_{\text{coh}}\equiv\left\langle \psi_{0}^{\prime}\right.\left|\psi_{0}\right\rangle$ is obtained by using Eq.~(\ref{coherentstate}),
\begin{eqnarray}
\mathcal{O}_{\text{coh}} =e^{\left(-\left|\psi_{0}^{\prime}\right|^{2}-\left|\psi_{0}\right|^{2}+2\psi_{0}^{\prime}\psi_{0}\right)/2},
\label{ovlp_psi}
\end{eqnarray}
which reads $\mathcal{O}_{\text{coh}}\simeq 1$ under Bogoliubov approximation $\psi_{0}^{\prime}=\sqrt{N_{0}^{\prime}/2}\simeq\sqrt{N/2}\simeq
\psi_{0}$.

\iffalse
In the Bogoliubov ground state, the number of particles in the $k=0$ state is
$N_{0}=\left\langle \Psi_{0}^{B}\left|\hat{b}_{0}^{\dagger}\hat{b}_{0}\right|\Psi_{0}^{B}\right\rangle$. By substituting Eq.~(\ref{bogo_ground_stat}) we have
\begin{eqnarray}
N_{0} & =\left\langle \psi_{0}\left|\hat{b}_{0}^{\dagger}\hat{b}_{0}\right|\psi_{0}\right\rangle =\left\langle \psi_{0}\left|\psi_{0}^{*}\psi_{0}\right|\psi_{0}\right\rangle =\left|\psi_{0}\right|^{2}
\label{N_0}
\end{eqnarray}
\fi

\subsubsection{Occupation of finite momentum states}
In the Bogoliubov ground state $\left|\Psi_{0}^{B}\right\rangle $, pairs of bosons are virtually excited to state with $k$ and $-k$.
The average number of virtually excited bosons with wave vector $k$ is obtained from the Bogoliubov transformation (\ref{transf}) and the definition of Bogoliubov ground state (\ref{psi_bogo_def}),
\begin{eqnarray}
n_{k} & =\left\langle \Psi_{0}^{B}\right|\hat{b}_{k}^{\dagger}\hat{b}_{k}\left|\Psi_{0}^{B}\right\rangle=\left|\nu\right|^{2}.
\label{nk}
\end{eqnarray}
We denote $N_{k\neq0}$ as the number of virtually excited particles, i.e. the number of particles in the state $\left|k\neq0\right\rangle$,
\begin{eqnarray}
N_{k\neq0}=2\sum_{k>0}n_{k}=2\sum_{k>0}\left|\nu\right|^{2}.
\label{N_k}
\end{eqnarray}
%where the factor $2$ comes from the fact that pairs of particles are excited into $\pm k$ states, so that the number of particles in state $\left|k\right\rangle $ is equal to that in state $\left|-k\right\rangle $. 

\subsubsection{Results}

\begin{figure}
	\centering\includegraphics[width=0.99\linewidth]{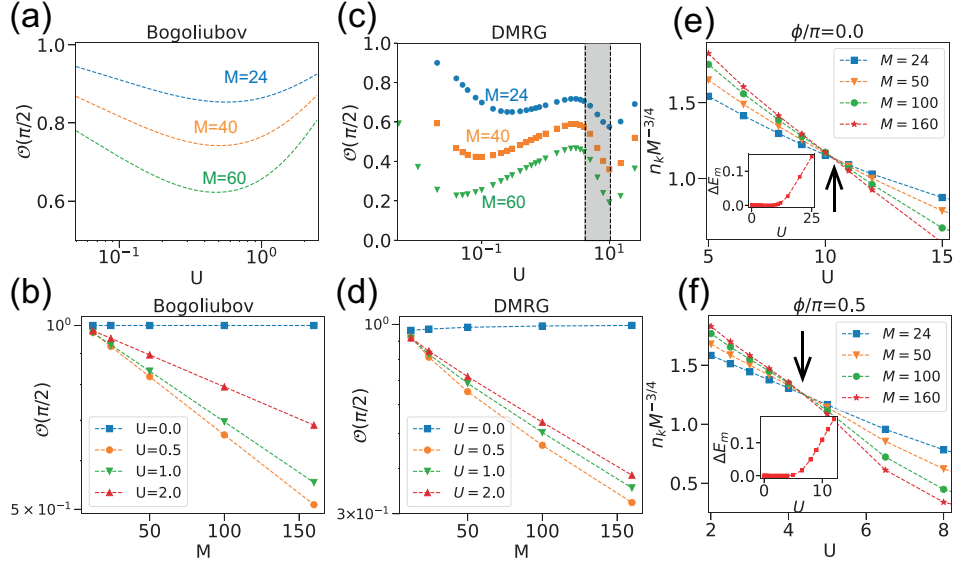}
	\caption{ (a) Semi-log plot of ground state overlaps $\mathcal{O}(\pi/2)$ from Bogoliubov theory as a function of interaction $U$, for different number of rungs $M$. (b) Semi-log plot of $\mathcal{O}(\pi/2)$ as a function of $M$ for different $U$. (c,d) Same plot as (a,b), but for DMRG simulations. Both (b) and (d) shows exponentially decay of the overlap with respect to $M$. (e) Scaling quasimomentum peak $n_k M^{-3/4}$ as a function of $U$ for different $M$. The crossing corresponds to BKT-transition points, which can be further confirmed by the mass gap shown in the inset.} 
	\label{fig_ovl}
\end{figure}

Now we apply the above expressions in our ladder system at $1/2$ filling with $J_{\perp}=2$.
We plot the analytic result for the overlap Eq.~(\ref{overlap}) for $M$-rung ladder with periodic boundary condition in Fig.~\ref{fig_ovl}(a), which shows qualitative agreement with the dip behavior in the weakly interacting regime from the DMRG simulations of \emph{finite} system with open boundary conditions [Fig.~\ref{fig_ovl}(c)]. Note that the DMRG results for the interacting regime have been divided into three regions. The beginning and the end of the grey shaded region are given by the BKT transition from a superfluid to a Mott insulator for $\phi=\pi/2$ and $\phi=0$, respectively. By extracting from the finite-size scaling of peaks in quasimomentum distribution $n^{\text{max}}_k M^{-3/4}$~\cite{1973Kosterlitz,2012Dhar,2016Greschner}, the crossing determines the BKT-transition points at $U_{c1}\approx10.4$ for $\phi=0$ and $U_{c2}\approx4.2$ for $\phi=\pi/2$ [Fig.~\ref{fig_ovl}(e,f)]. 
Overall, both the analytic and numerical results show that the overlaps decay exponentially with the system size for finite $U$, and approach $1$ for the non-interacting case~[Fig.~\ref{fig_ovl}(b,d)].

\begin{figure}
	\centering\includegraphics[width=0.75\linewidth]{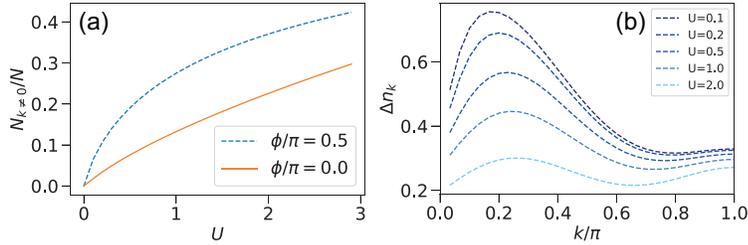}
	\caption{(a) Number of particles with non-zero quasi-momentum for $\phi=0$~(solid line) and $\phi=\pi/2$~(dashed line), scaled with total particle number $N$. (b) Difference of non-zero $k$ mode occupation between $\phi=\pi/2$ and $\phi=0$. Here we choose the number of rungs $M=60$, number of particles $N=60$, and $J_{\perp}=2$.}
	\label{fig_Ek}
\end{figure}

To understand the dip in the weakly interacting regime, we plot the average number of particles with non-zero quasi momentum $N_{k\neq0}$ according to Eq.~(\ref{N_k}), and the relative difference in the occupation of non-zero k-modes $\Delta n_{k}=\frac{n_{k}(\pi/2)-n_{k}(0)}{n_{k}(\pi/2)+n_{k}(0)}$ between $\phi=\pi/2$ and $\phi=0$ in Fig.~\ref{fig_Ek}(a) and (b), respectively.
We can observe that when switching on the interactions, the excited quasi momentum modes become occupied much faster in the presence of magnetic flux. This is related to the fact that the single-particle dispersion relation $E_-(k)$ acquires a larger effective mass with increasing flux [see Fig. 4(a) in the main text]. As a result, the momentum modes become occupied rather differently for both fluxes when U is switched on, as can be seen from Fig.~\ref{fig_Ek}(b). The slight increase of the overlap for even larger $U$ can then be explained by the fact that the relative differences in the momentum distributions for both fluxes become smaller again.

\newpage
\section*{References}

\bibliographystyle{iopart-num}
\bibliography{mybib}

\end{document}